# Do Ethical AI Principles Matter to Users?
# A Large-Scale Analysis of User Sentiment and Satisfaction

Stefan Pasch[1]  Min Chul Cha[2]

## Abstract

As AI systems become increasingly embedded in organizational workflows and consumer applications, ethical principles such as fairness, transparency, and robustness have been widely endorsed in policy and industry guidelines. However, there is still scarce empirical evidence on whether these principles are recognized, valued, or impactful from the perspective of users. This study investigates the link between ethical AI and user satisfaction by analyzing over 100,000 user reviews of AI products from G2.com. Using transformer-based language models, we measure sentiment across seven ethical dimensions defined by the EU Ethics Guidelines for Trustworthy AI. Our findings show that all seven dimensions are positively associated with user satisfaction. Yet, this relationship varies systematically across user and product types. Technical users and reviewers of AI development platforms more frequently discuss system-level concerns (e.g., transparency, data governance), while non-technical users and reviewers of end-user applications emphasize human-centric dimensions (e.g., human agency, societal well-being). Moreover, the association between ethical AI and user satisfaction is significantly stronger for non-technical users and end-user applications across all dimensions. Our results highlight the importance of ethical AI design from the user's perspective and underscore the need to account for contextual differences across user roles and product types.

## 1. Introduction

With the increasing integration of artificial intelligence (AI) systems into organizational workflows and daily life, the ethical and societal implications of these technologies have become a focal point of academic, policy, and industry debates (Floridi & Cowls, 2022; Brundage et al., 2020; Fjeld et al., 2020). In response, numerous initiatives have been introduced to encourage the ethical and responsible use of AI technologies. Among the most prominent are the European Commission's Ethics Guidelines for Trustworthy AI (AI HLEG, 2019), and the OECD AI Principles (2019), alongside a growing landscape of national strategies and corporate frameworks (Jobin et al., 2019; Fjeld et al., 2020). Despite differences in scope and emphasis, these efforts have converged around a shared set of high-level principles; typically including transparency, fairness, accountability, technical robustness, human oversight, and broader societal benefit.

While these frameworks have helped consolidate a shared vocabulary around ethical AI, they are largely top-down in nature: typically developed by expert committees, policymakers, or institutional bodies, rather than through engagement with end users or practitioners (Mittelstadt, 2019; Jobin et al., 2019). As such, their legitimacy rests in part on the assumption that these principles are not only normatively sound but also meaningful and relevant in applied contexts. Yet empirical evidence remains scarce on whether users recognize, value, or are affected by these ethical dimensions in their actual interactions with AI systems (Fernholz et al., 2024; Sigfrids et al., 2023). Without such user-centered validation, ethical AI risks becoming a compliance exercise: procedurally implemented but perceptually absent – decoupled from the expectations, experiences, and satisfaction of those who interact with AI in practice.

---

[1] Division of Social Science & AI, Hankuk University of Foreign Studies: stefan.pasch@outlook.com
[2] Corresponding Author. Division of Social Science & AI, Hankuk University of Foreign Studies: minchulcha@hufs.ac.kr

Understanding how users perceive and respond to ethical principles is not only important for policymakers, who must ensure that governance frameworks are socially grounded and widely adopted, but also for the fields of Human-AI Interaction (HAI) and Human-Computer Interaction (HCI), where trust, usability, and alignment with user values are central concerns. If users are unaware of, misunderstand, or are indifferent to ethical dimensions, even well-intentioned frameworks may fail to foster trust, engagement, or satisfaction. Conversely, ethical design elements that are perceptible and meaningful at the interface level may enhance legitimacy and user acceptance, especially in high-stakes or sensitive contexts.

To investigate how ethical principles shape user experiences, we draw on the EU Ethics Guidelines for Trustworthy AI (AI HLEG, 2019), a widely cited framework that has informed both policy and academic debates. These guidelines identify seven dimensions of trustworthy AI: human agency and oversight, technical robustness and safety, privacy and data governance, transparency, diversity and fairness, societal and environmental well-being, and accountability. The framework has gained significant traction due to its comprehensiveness and normative influence and has been validated in recent academic work as capturing core themes across the ethical AI landscape (Floridi & Cowls, 2022; Papagiannidis et al., 2023).

Building on this foundation, we address the following research question:

***RQ1: How do the ethical AI dimensions (as defined by the EU AI Ethics Guidelines) relate to user satisfaction in real-world AI interactions?***

While this question addresses the overall relationship between ethical AI and satisfaction, it is unlikely that all users engage with ethical principles in the same way. Both HCI theory and prior empirical research suggest that users' backgrounds and the type of AI system they interact with can shape how ethical concerns are perceived, understood, and valued (Ko et al., 2006; Carroll, 1997; Laato et al., 2022).

This is particularly important in the context of AI, where some products focus on the development and deployment of AI (e.g. MLOps or Data Science platforms) while in other products AI is integrated in the end-product (e.g. AI-based marketing tools). Similarly, users with different levels of technical background will interact with AI product: Technical users, such as data scientists, who are familiar with the system logic of AI products and interested in technical details, and non-technical users who may focus more on usability, outcomes, and broader societal impacts and less on system-specific details.

Despite these differences, ethical AI guidelines, such as those developed by the EU, typically assume a one-size-fits-all approach, without tailoring principles to different user roles or product contexts. As a result, it remains unclear whether these ethical dimensions are equally recognized or valued across use cases. Specifically, differences in user background and product type may shape both (a) the extent to which particular ethical principles are salient to users and correspondingly discussed in reviews; and (b) how strongly those principles influence satisfaction. This leads to the following research questions:

***RQ2: How does the degree to which ethical AI dimensions are mentioned vary by users' professional background and the type of AI product reviewed?***

***RQ3: Do users' professional background and the type of AI product moderate the relationship between ethical AI dimensions and user satisfaction?***

To empirically examine these questions, we draw on a large-scale dataset of over 100,000 user reviews of AI products from the software review platform G2. We develop a novel approach to measure ethical AI dimensions based on the free-text content of these reviews. Specifically, we label the sentiment of a subset of reviews along the seven dimensions defined in the EU Ethics Guidelines for Trustworthy AI using zero-shot classifications. We then use these annotations to fine-tune transformer-based language models – state-of-the-art tools for domain-specific text classification – which allow us to predict sentiment scores for each ethical dimension across the full dataset.

Our results suggest that all seven dimensions of ethical AI from the EU Guidelines are positively associated with user satisfaction, indicating that ethical principles are not only normative ideals but also align with users' experiential evaluations. However, we observe systematic differences in how frequently different dimensions are discussed. Technical users and reviewers of AI development platforms are more likely to mention system-level concerns, such as technical robustness, transparency, and data governance, while non-technical users and reviewers of end-user applications more often emphasize human-oriented aspects like fairness, human agency, and societal impact.

Interestingly, we find a different pattern when examining how ethical AI dimensions influence satisfaction: across all dimensions, the effect on satisfaction is consistently weaker for technical users and reviews of AI development platforms, suggesting that ethical considerations may be more consequential for those with less control or technical expertise.

Taken together, this paper makes three main contributions. First, we present a novel and scalable method for measuring user sentiment on ethical AI dimensions by applying transformer-based language models to a large corpus of real-world product reviews. This approach enables fine-grained, dimension-specific analysis of how users evaluate AI products along ethical lines. Second, we provide empirical evidence that the seven ethical principles outlined in widely used frameworks, such as the EU Ethics Guidelines for Trustworthy AI, are not just normative ideals, but correlate positively with user satisfaction. This highlights the experiential relevance of ethical AI for users in practice. Third, we show that user background (technical vs. non-technical) and product type (AI development platforms vs. end-user applications) systematically shape both the ethical concerns users discuss and the strength of the ethics–satisfaction relationship. These findings underscore the importance of contextualizing ethical AI not only in terms of system design but also in terms of who interacts with the system, and how.

## 2. Literature Review

### 2.1. Ethical AI and the EU's Ethics Guidelines for Trustworthy AI

The concept of ethical AI has become central to global efforts to ensure that artificial intelligence systems align with human values and uphold societal norms. Increasingly, ethical AI is understood to encompass both normative principles, such as fairness, autonomy, and accountability, and the enabling mechanisms necessary to realize them in practice, including transparency, data governance, and technical robustness (Floridi & Cowls, 2022; Fjeld et al., 2020; Jobin et al., 2019). Ethical AI, in this view, is not reducible to abstract ideals but must be grounded in concrete design and oversight strategies that minimize harm and safeguard fundamental rights.

In response, recent years have seen the emergence of numerous frameworks on ethical AI and related concepts, such as *responsible AI*, *trustworthy AI*, and *AI governance*, developed by institutional actors, academic initiatives, and private-sector stakeholders. These frameworks typically aim to define key ethical principles, identify risk areas, and offer practical guidance for the design, deployment, and oversight of AI systems. Examples include the European Commission's Ethics Guidelines for Trustworthy AI (AI HLEG, 2019), the OECD AI Principles (2019), and the UNESCO Recommendation on the Ethics of Artificial Intelligence (2021), as well as voluntary codes of conduct issued by major technology firms (Fjeld et al., 2020). While these frameworks differ in emphasis and scope, comparative reviews have found substantial convergence around a core set of values, including fairness, accountability, transparency, privacy, and safety (Jobin et al., 2019; Fjeld et al., 2020).

In this paper, we operationalize ethical AI principles using the European Commission's Ethics Guidelines for Trustworthy AI (AI HLEG, 2019). This choice is motivated by three main considerations. First, the framework holds strong policy relevance. As one of the earliest and most comprehensive governmental initiatives in AI ethics, it has significantly shaped European AI governance, including the development of the EU Artificial Intelligence Act, which represents the first major legislative proposal on AI. The EU's approach is widely regarded as a global benchmark and has contributed to what scholars describe as the "Brussels Effect" in technology regulation. Its influence extends beyond Europe, informing governance efforts in countries around the world, including OECD members such as Canada and Japan. (Siegmann & Anderljung, 2022). Its status as an official guideline issued by a democratic, supranational body lends it both normative legitimacy and regulatory weight.

Second, the EU Guidelines demonstrate broad cross-sector uptake. They have been applied not only by EU institutions, but also by companies, public bodies, and industry associations to guide ethical impact assessments, procurement criteria, and internal governance processes (Morley et al., 2021). The accompanying Assessment List for Trustworthy AI (ALTAI) was piloted by more than 350 organizations and continues to be cited in both academic research and applied policy contexts (European Commission, 2020). This reflects its practical traction and relevance across diverse implementation settings.

Third, the framework's seven dimensions, including fairness, transparency, human agency, and technical robustness, are closely aligned with theoretical and empirical syntheses in the AI ethics literature. Papagiannidis et al. (2023), in a review of 48 studies on responsible AI, find that dominant concerns in the field can be meaningfully categorized under the seven dimensions introduced by the EU Guidelines. Furthermore, Floridi and Cowls (2022) argue that all of the five core ethical principles they identified are captured within the EU framework. This supports its conceptual completeness and its suitability for structuring empirical analysis.

Table 1. Dimensions of the EU Ethics Guidelines for Trustworthy AI

| Dimension | Description | Type |
| --- | --- | --- |
| Human Agency & Oversight | AI should support human autonomy and allow for meaningful human oversight. | Human-Oriented |
| Diversity, Non-discrimination & Fairness | AI should promote fairness and avoid unjust bias and exclusion. | Human-Oriented |
| Societal & Environmental Well-being | AI should benefit society broadly and promote environmental sustainability. | Human-Oriented |
| Accountability | Clear responsibility should be assigned for AI outcomes, with redress mechanisms in place. | Human-Oriented |
| Technical Robustness & Safety | AI systems should be secure, reliable, and resilient to harm or failure. | System-Oriented |
| Privacy & Data Governance | AI must respect privacy and ensure quality and legitimate access to data. | System-Oriented |
| Transparency | AI processes should be explainable, traceable, and clearly communicated. | System-Oriented |

The *Ethics Guidelines for Trustworthy AI* articulate seven dimensions (see Table 1 for an overview) that together define what constitutes ethically aligned AI. These dimensions not only structure the EU's governance approach but also reflect long-standing concerns in the academic literature on AI ethics and responsible technology design (Floridi & Cowls, 2022; Jobin et al., 2019).

To structure our analysis, we group the EU's seven dimensions into two broad categories. This reflects a conceptual distinction increasingly recognized in AI ethics scholarship: (i) *human-oriented values*, which refer to ethical commitments centered on human dignity, justice, and human roles and responsibilities; and (ii) *system-level features*, which refer to the technical mechanisms and system properties necessary to operationalize AI safely and ethically. The distinction between human-oriented values and system-level features is well-supported in both academic literature and policy frameworks. Scholars such as Floridi and Cowls (2019) emphasize that ethical principles like fairness, autonomy, and justice must be accompanied by operational mechanisms, such as transparency, technical robustness, and data governance, that enable their practical realization. This dual structure reflects a broader consensus in the field (e.g., Jobin et al., 2019; Fjeld et al., 2020).

Among the seven ethical dimensions defined by the EU Guidelines, four can be grouped under human-oriented values, those that focus on people, justice, and society. *Human agency and oversight* emphasizes autonomy, informed decision-making, and meaningful human control. *Diversity, non-discrimination, and fairness* focuses on inclusive design and the mitigation of unjust bias. *Societal and environmental well-being* addresses the broader social and ecological impacts of AI, aligning with calls for sustainable and socially beneficial innovation. *Accountability* pertains to institutional responsibility, mechanisms for redress, and the ability to challenge or contest AI-driven decisions.

The remaining three dimensions fall under system-oriented features, referring to the technical and procedural mechanisms necessary to implement ethical values. *Technical robustness and safety* covers system reliability, resilience, and security, which are central to discussions of safe and trustworthy AI. *Privacy and data governance* refers to the protection, quality, and integrity of data, along with appropriate mechanisms for access and control. *Transparency* involves traceability, explainability, and clear communication about system functionality, reflecting long-standing debates on algorithmic interpretability and epistemic responsibility.

While *transparency* and *accountability* can both be seen as bridging dimensions – connecting system design with human oversight – we treat them differently here based on their primary locus of concern. Transparency is classified as a system-oriented feature due to its emphasis on internal mechanisms such as explainability and traceability, whereas accountability is treated as a human-oriented value, as it deals with governance structures, responsibility assignment, and access to redress.

In sum, the EU's Ethics Guidelines for Trustworthy AI encompass both high-level ethical human-oriented values and system-level properties, offering a comprehensive view of what constitutes ethically aligned AI. While designed primarily for governance and development settings, these dimensions may also inform how users perceive AI systems in applied contexts.

In the next section, we examine whether and how these ethical considerations relate to user satisfaction.

## 2.2. Ethical AI and User Satisfaction

Ethical principles of AI development and deployment have been predominantly addressed at the policy level, focusing on governance, regulation, and high-level design criteria. Correspondingly, empirical evidence on how such ethical principles, such as those articulated in the EU Ethics Guidelines for Trustworthy AI, affect the experience of users remains scarce (Fernholz et al., 2024; Sigfrids et al., 2023). However, despite their policy origins, these guidelines are ultimately intended to shape the design and deployment of AI systems in ways that are meaningful for end-users. As suggested by related literature in human-computer interaction (HCI), the successful implementation of ethical principles should not only satisfy normative criteria but also translate into enhanced user experience and satisfaction.

A central mechanism linking ethically aligned AI design to user satisfaction is trust – the belief that a system will behave reliably, transparently, and in the user's best interest (Rai, 2020; Shneiderman, 2020). Trust can be fostered through system-level features such as technical robustness, predictability, and explainability, which help users form accurate mental models of system behavior and reduce uncertainty (Ehsan et al., 2024; Gupta et al. 2019). At the same time, human-oriented values like fairness, accountability, and respect for autonomy contribute to trust by signaling that the system's operation aligns with ethical norms and societal expectations (Binns, 2018; Wieringa, 2020). Trust, in turn, has been linked to overall satisfaction with AI systems (Choudhury & Shamszare, 2023).

Satisfaction with AI systems is also shaped by how easily users can interact with them and exert meaningful control. System-level features such as explainability, responsiveness, and consistent performance reduce cognitive load and support smoother interactions – key contributors to usability and user confidence (Amershi et al., 2019; Yang et al., 2019). At the same time, human-oriented principles like human agency and oversight enhance users' sense of control, especially when systems provide mechanisms for intervention, feedback, or override. These capabilities, often supported by interpretability and auditability, enable users to feel more autonomous and less alienated in their interactions. Consistent with technology acceptance models (Davis, 1989; Venkatesh et al., 2003), such designs promote satisfaction by making systems not only easier to use but also more responsive to human judgment and responsibility (Shneiderman, 2020; Amershi et al., 2019).

Another important channel is value alignment. Users are more satisfied with technologies that reflect shared social and ethical commitments, such as fairness, inclusivity, sustainability, and privacy (Kieslich et al., 2021; Knijnenburg & Kobsa, 2013). These values, central to the EU's human-oriented dimensions, resonate when operationalized through system-level features like bias mitigation, privacy-preserving architectures, and participatory design. When users perceive that AI systems reflect and uphold these values, they report greater emotional comfort and moral satisfaction (Yang & Lee, 2024).

Closely related is the perception of procedural fairness and accountability. Users not only value alignment with ethical ideals but also want to see that systems treat people equitably and offer clear pathways for contesting harmful outcomes. While fairness and accountability are primarily human-oriented concerns, they are enabled by system-level capabilities like auditability, documentation, and redress mechanisms. When such structures are visible and credible, users are more likely to view the system as legitimate, a critical ingredient for sustained satisfaction (Binns, 2018; Holstein et al., 2019).

Taken together, these pathways suggest that ethically aligned AI design – encompassing both human-oriented values and system-level features – can meaningfully shape how users perceive and evaluate AI systems. We therefore hypothesize:

***H1: Sentiment on ethical AI dimensions is positively associated with user satisfaction.***

### 2.3. Discussion of Ethical AI Dimensions and Job Role and Product Type

While we hypothesize that ethical AI dimensions are positively linked with user satisfaction, the extent to which users engage with these dimensions is likely to vary. In particular, two contextual factors may systematically shape how users perceive and prioritize different aspects of ethical AI: the professional background of the user and the type of AI product being evaluated. These factors reflect deeper differences in how users relate to AI systems; whether they interact with AI as a configurable technical system or as a source of outputs, decisions, or content. We first focus on how these factors influence the discussion level of ethical dimensions – that is, the degree to which different principles are explicitly emphasized in user feedback.

The Human-Computer Interaction (HCI) literature has long recognized that users' roles, domain knowledge, and task demands shape how they interact with and evaluate technology (Grudin, 1991; Fischer, 2001; Ko et al., 2006). For example, Grudin (1991) distinguishes between technical users, such as developers or engineers, who engage directly with system internals and emphasize attributes like reliability, transparency, and system integrity; and non-technical users, who focus more on usability, intuitiveness, and how seamlessly technology integrates into their workflows.

Subsequent research has reinforced this distinction. Technical professionals often prioritize system-level qualities because their work involves configuring, analyzing, or improving technical systems (Ko et al., 2006; Blackwell, 2002). In contrast, non-technical users tend to focus on whether a system is easy to use and whether it helps them achieve their practical goals, rather than on how the system functions internally (Carroll, 1997; Davis, 1989). This difference reflects broader distinctions between system-centric evaluation and interaction- or outcome-oriented evaluation in HCI (Gasson, 2003; Díaz et al., 2008).

Recent studies in Human-AI Interaction (HAI) echo this pattern, showing that technical users are more attentive to system performance, explainability, and reliability, whereas non-technical users emphasize interaction quality and personalization (Pasch & Ha, 2025; Jiang et al., 2024). These distinctions closely mirror the framework of ethical AI used in this study: system-level features (e.g., technical robustness, data governance, transparency) versus human-oriented values (e.g., fairness, accountability, human agency). More broadly, this distinction reflects where users are situated in the AI lifecycle: technical users are more likely to interact with AI as a configurable system, whereas non-technical users more often encounter AI as a decision-making or content-producing product. This distinction between system-facing and outcome-facing roles is rarely made explicit in unified AI ethics frameworks, but it meaningfully shapes which ethical dimensions are most salient to users in practice.

Accordingly, we expect technical users to reference system-level dimensions of ethical AI more frequently. By contrast, non-technical users may place greater emphasis on human-oriented ethical concerns, as they are more directly exposed to the practical and societal outcomes of AI deployment. This includes concerns closely tied to interaction quality, such as human agency, as well as broader normative values such as fairness and societal well-being, which reflect whether systems behave in ways that align with social expectations and equitable treatment.

In addition to user background, the type of AI product itself may shape how ethical dimensions are perceived and discussed. We distinguish between AI development platforms, such as data science platforms, and end-user applications, such as AI-based marketing, sales, or HR tools. This distinction reflects differences in the intended use, depth of user interaction with AI functionality, and the types of concerns users are likely to encounter.

AI development platforms are typically used by engineers, data scientists, and other specialists who engage deeply with the configuration, monitoring, and optimization of AI systems. These tools expose users to aspects of system behavior, including model training, deployment, and error handling. As a result, evaluations of these platforms are likely to center on system-level features of ethical AI, such as robustness, transparency, and data governance, which directly affect system performance, reliability, and developer trust (Zhou et al., 2020; Breck et al., 2017).

In contrast, end-user applications abstract away from technical details and focus instead on delivering AI-driven outcomes, such as content generation, personalization, or automated decision support. Users of these products interact with AI primarily through its outputs and interface, not through its inner workings. Consequently, discussions of ethical AI in this context are more likely to revolve around human-oriented values, such as fairness, inclusivity, or the perceived societal impact of algorithmic decisions (Laato et al., 2022). Users may express concern about biased recommendations, lack of control, or the appropriateness of AI decisions in socially sensitive domains.

Accordingly, we expect system-level dimensions to be more frequently discussed in evaluations of AI development platforms, and human-oriented concerns to be more prominent in feedback on end-user applications. This reflects the different focal points of each product type: while technical platforms expose users to internal system functioning, end-user tools foreground practical outcomes and social impact.

*H2a: Discussion of ethical AI dimensions differs by user role, such that system-level dimensions are more frequently discussed by technical users, while human-oriented dimensions are more frequently discussed by non-technical users.*

*H2b: Discussion of ethical AI dimensions differs by product type, such that system-level dimensions are more frequently discussed in reviews of AI development platforms, while human-oriented dimensions are more frequently discussed in reviews of end-user applications.*

## 2.4. Moderating Role of Job Role and Product Type on the Ethical AI-User Satisfaction Link

While user background and product type may shape which ethical AI dimensions are emphasized in feedback, they may also influence how these dimensions translate into user satisfaction. In other words, beyond shaping discussion levels, contextual factors could moderate the strength of the relationship between ethical AI principles and perceived system value.

This distinction is consistent with classic models in HCI and technology acceptance that emphasize the contextual nature of satisfaction. As Venkatesh and Davis (2000) note, the factors driving satisfaction are not fixed, but vary based on user roles, domain knowledge, and task environments. Accordingly, we hypothesize that ethical AI dimensions do not uniformly affect satisfaction, but instead interact with user and product characteristics in systematic ways.

One line of argument is that the same considerations that influence which ethical AI dimensions users emphasize are also likely to affect which dimensions affect the ethical AI-satisfaction link. Since system-level properties directly impact technical users' ability to evaluate, debug, and optimize AI systems, core components of their daily workflows, these features may have a stronger effect on satisfaction by more directly supporting users' goals and task performance (Ko et al., 2006; Sculley et al., 2015). Similarly, users of developer-oriented tools, such as MLOps platforms, regularly engage with system internals and are often responsible for ensuring technical robustness, data governance, and system reliability. As a result, their satisfaction may be more strongly shaped by system-level ethical AI dimensions, as these directly affect the functionality, reliability, and control required in their professional use of AI systems.

In contrast, non-technical users and users of end-user applications interact with AI through its outputs and interfaces, making their experience more closely tied to interaction quality, usability, and alignment with social values (Carroll, 1997; Davis, 1989; Pasch & Ha, 2025). For this group, satisfaction tends to depend on whether the system supports autonomy, fairness, inclusivity, and social legitimacy, dimensions central to human-oriented ethical principles (Laato et al., 2022).

***H3a: The relationship between ethical AI dimensions and user satisfaction is moderated by user role, such that system-level dimensions are more strongly associated with satisfaction among technical users, whereas human-oriented dimensions are more strongly associated with satisfaction among non-technical users.***

***H3b: The relationship between ethical AI dimensions and user satisfaction is moderated by product type, such that system-level dimensions are more strongly associated with satisfaction for AI development platforms, whereas human-oriented dimensions are more strongly associated with satisfaction for end-user applications.***

While the argument above suggests that alignment between user roles and ethical AI dimensions may strengthen their impact on satisfaction, an alternative perspective focuses less on alignment and more on the *nature of users' relationship to AI systems*. In particular, not only do technical and non-technical users differ in their task focus, but also in how they relate to AI across different layers of the AI lifecycle. Technical users interact with AI systems at the level of configuration, monitoring, or optimization – closer to their inner workings. In contrast, non-technical users engage with AI systems primarily through their outputs, such as content, recommendations, or decisions, and do not shape the system's behavior directly.

For technical users, AI tools are often experienced as systems to build, test, or manage. Engineers and data scientists engage with AI systems primarily to develop, deploy, and optimize machine learning models or AI-driven applications. Ethical shortcomings, such as a lack of fairness, unclear outputs, or brittle behavior, are not seen as final product failures, but as technical constraints to be managed or improved upon. These users typically possess the knowledge and agency to work around or fix such issues (Ko et al., 2006; Sculley et al., 2015), which gives them a higher *perceived control* over the technology (Beaudry & Pinsonneault, 2005). This perspective reflects longstanding observations in software engineering research, where system flaws are frequently treated by developers as solvable challenges rather than failure (Gasser, 1986; Nandhakumar & Avison, 1999). In terms of satisfaction, this perceived controllability and expectation of iteration reduces the emotional weight of ethical or technical flaws. Even system-level issues like robustness or explainability, while salient, may not meaningfully diminish satisfaction for these users, who see them as *fixable* rather than *fatal* problems.

By contrast, non-technical users typically encounter AI through its outputs – recommendations, classifications, or content. Here, the AI system is not a configurable tool but a finished product. As a result, ethical lapses such as biased decisions or a lack of transparency are interpreted as *failures of product quality*, not just areas for improvement. These users often lack the means to diagnose or remedy problems, making them more vulnerable to trust erosion and dissatisfaction (Laato et al., 2022). From the perspective of *expectancy-disconfirmation theory* (Oliver, 1980), non-technical users come with expectations of reliability, fairness, and usability. When these are not met, disconfirmation is greater, and satisfaction declines more sharply. Conversely, technical users often view such imperfections as expected constraints in an evolving system. Their familiarity with system internals and capacity to address issues enables a more tolerant, problem-solving mindset, which can reduce the impact of ethical shortcomings on satisfaction.

A similar pattern holds at the level of product type. AI development platforms, such as MLOps tools and data science environments, are designed for configuring, training, and deploying machine learning models. Users of these systems engage directly with algorithmic internals, performance metrics, and infrastructure components. As a result, ethical concerns like fairness or transparency may be noted, but they are often treated as technical challenges to be resolved within the development lifecycle. By contrast, end-user applications deliver AI-driven outputs, such as content, recommendations, or classifications, directly to business users or consumers. In these settings, users typically have no influence over how the system operates, making ethical failures more visible, less controllable, and more damaging to user trust and satisfaction.

*H4a: The relationship between ethical AI dimensions and user satisfaction is moderated by user role, such that this relationship is weaker for technical users than for non-technical users for all dimensions of ethical AI.*

*H4b: The relationship between ethical AI dimensions and user satisfaction is moderated by product type, such that this relationship is weaker for AI development platforms than for end-user applications for all dimensions of ethical AI.*

## 3. Methodology

Figure 1 provides a visual overview of our methodological pipeline. We begin by applying zero-shot classification to a sample of 3,000 user reviews, labeling sentiment across the seven ethical AI dimensions defined in the EU Ethics Guidelines. These labeled data are used to fine-tune a transformer-based classifier for dimension-specific sentiment detection. The resulting models are then applied to the full dataset of over 100,000 reviews to generate structured sentiment scores for each ethical dimension. Finally, we analyze how these scores relate to user satisfaction ratings, allowing us to assess the relevance of ethical AI principles in real-world user experiences.

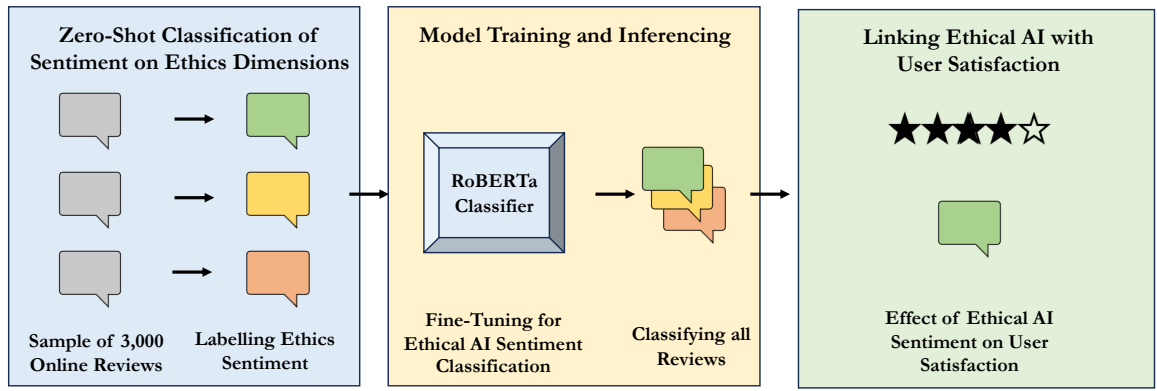

**Figure 1.** Overview of Methodological Pipeline

### 3.1 Data

This study draws on user-generated reviews from G2.com, a prominent platform for evaluating business-to-business (B2B) software. Unlike general consumer review sites such as Amazon or Trustpilot, G2 focuses specifically on enterprise and workplace technologies, making it particularly suitable for understanding how AI systems are experienced in professional contexts (G2, 2024; Kevans, 2023). The platform includes reviews for both technical tools used by developers and infrastructure teams, as well as end-user applications adopted by broader business functions. This diversity enables a comparative analysis of AI perception across user roles and product types.

We sourced data from G2's "Artificial Intelligence" category, which covers a broad range of products, including MLOps platforms, chatbots, AI writing assistants, recommendation engines, and intelligent automation tools. To ensure a sufficient volume of data for each product, we restricted our sample to those with at least 50 reviews. The final dataset consists of 249 AI products and 108,998 individual reviews.

Each review consists of two open-ended response fields: one asking what the user likes about the product, and another asking what they dislike. To retain this polarity in our analysis, we combined the two text segments into a single review body, prefixing each section with clear markers – "Like:" and "Dislike:" respectively. This formatting helps natural language models distinguish between positively and negatively framed content, which is important for accurately capturing sentiment toward specific ethical AI dimensions.

In addition to the review text, the dataset includes structured metadata for each entry. This includes the job title of the reviewer, which we use to infer whether the user holds a technical or non-technical role, as well as information on the product's functional category. Each review also contains a star rating (ranging from 1 to 5), which serves as a quantitative measure of user satisfaction. Together, these components offer a rich foundation for analyzing how ethical AI concerns are discussed and how they relate to user satisfaction across different contexts.

## 3.2 Zero-Shot Classifications for Labeling

Table 2. Distribution of Ethical AI Dimensions (in %)

| Dataset | Random Sample | | | Corrected Sample | | | Predictions on Full Dataset | | |
|---|---|---|---|---|---|---|---|---|---|
| **Dimension/ Sentiment Label** | Not Disc. | Pos. | Neg. | Not Disc. | Pos. | Neg. | Not Disc. | Pos. | Neg. |
| Human Agency & Oversight | 55.2 | 42.7 | 1.5 | 54.9 | 42.4 | 2.6 | 58.1 | 40.4 | 1.5 |
| Diversity & Fairness | 94.0 | 1.9 | 4.0 | 83.2 | 6.2 | 10.5 | 91.5 | 2.6 | 5.8 |
| Societal & Env. Well-being | 96.4 | 2.7 | 0.8 | 87.9 | 9.4 | 2.6 | 95.9 | 3.3 | 0.6 |
| Accountability | 87.7 | 3.7 | 8.7 | | | | 89.7 | 3.4 | 6.9 |
| Technical Robustness & Safety | 32.7 | 43.2 | 24.1 | | | | 24.7 | 51.9 | 23.3 |
| Privacy & Data Governance | 93.7 | 1.3 | 4.9 | 77.8 | 8.8 | 13.4 | 91.6 | 1.6 | 6.8 |
| Transparency | 78.8 | 13.1 | 8.1 | | | | 80.2 | 13.4 | 6.2 |

The first step in classifying sentiment toward ethical AI principles was to construct a labeled dataset that could be used to train supervised models for large-scale analysis. To do so, we applied zero-shot classification using *LLaMA 3.3 70B*, a state-of-the-art open-source generative language model. Prior research has demonstrated that large language models (LLMs) such as LLaMA can produce high-quality domain-specific annotations and, in some cases, outperform human crowdworkers (Chae & Davidson, 2023; Törnberg, 2023; Pasch & Cutura, 2024). Moreover, recent studies have validated the use of zero-shot LLMs for concept extraction in HCI and UX research (Pasch et al., 2025).

We randomly sampled 3,000 reviews from the full dataset and instructed the model to classify each review with respect to the seven ethical AI dimensions outlined by the EU High-Level Expert Group on AI (AI HLEG, 2019). For each dimension, the model was prompted with its definition and asked to assign one of four sentiment labels: *positive*, *negative*, *neutral*, or *not discussed*. Since neutral classifications were exceedingly rare (below 1% of all classifications), we combined *neutral* and *not discussed* into a single "not discussed" category.

The initial analysis revealed considerable variation in how frequently each dimension was discussed. Technical robustness and transparency, for example, were frequently mentioned with both positive and negative sentiment, whereas dimensions such as diversity & fairness or societal and environmental well-being were skewed toward omission. To correct for this imbalance and ensure the training set included sufficient examples of all classes, we applied a targeted sampling approach for underrepresented dimensions; specifically for the dimensions human agency and oversight, privacy and data governance, diversity and fairness, and societal and environmental well-being.

We used a BERT-based semantic similarity model (Mulyar, 2024) to calculate the textual similarity between each underrepresented ethical dimension's definition and every review in the full corpus. For those dimensions we selected the top 1,000 reviews with the highest similarity scores – those most likely to mention the concept – and subjected them to zero-shot classification using the same LLM pipeline. This targeted oversampling increased the representation of rare but important classes (e.g., negative sentiment on societal and environmental well-being) while preserving the overall distribution for well-represented dimensions.

The resulting corrected dataset was then used for model training and evaluation. Notably, this oversampling did not lead to a skewed distribution in the final predictions on the full dataset. Instead, the model outputs closely mirrored the distribution observed in the initial random sample – rather than the corrected sample – ensuring that the large-scale predictions retained the original characteristics of the data (see Table 2).

## 3.3. Fine-Tuning RoBERTa for Text Classification

In the next step, the labeled dataset was split into training (70%), validation (10%), and test (20%) sets, a common approach in supervised machine learning to ensure robust model evaluation. We then fine-tuned a separate classifier for each ethical AI dimension using RoBERTa-Large (Liu et al., 2019). RoBERTa-Large is a transformer-based language model that is widely regarded as state-of-the-art for domain-specific text classification tasks due to its strong ability to capture contextual and semantic nuance (González-Carvajal & Garrido-Merchán, 2020; Koch & Pasch, 2023). This makes it particularly well-suited for identifying sentiment toward abstract and technical concepts, such as ethical AI dimensions.

For the training process we relied on an established setup for fine-tuning RoBERTa for text classification (Pasch et al. 2025): Each RoBERTa-Large model was fine-tuned for 12 epochs using a learning rate of 1e-5 and a batch size of 8. To prevent overfitting, we applied a weight decay of 0.01 throughout the optimization process. Model performance was monitored on the validation set at regular intervals, with checkpoints evaluated every three epochs. The best-performing model for each ethical AI dimension was selected based on the highest weighted F1 score on the validation set at these checkpoints.

Training was conducted using the AdamW optimizer, which is well-suited for large transformer architectures due to its combination of adaptive learning rates and integrated weight regularization. To promote training stability and efficient convergence, we employed a cosine learning rate schedule with a linear warm-up phase. During the initial 10% of training steps, the learning rate increased linearly before gradually tapering off in a cosine pattern, helping to avoid abrupt gradient updates early in training and ensuring smooth optimization.

Table 3 presents the classification performance of each fine-tuned RoBERTa model on the holdout test set across the seven ethical AI dimensions. The resulting F1 scores and accuracy metrics range between 74% and 92%, aligning with established benchmarks for domain-specific text classification tasks (Koch & Pasch, 2023; Alturayeif et al., 2023), and demonstrating the effectiveness of transformer-based models for identifying sentiment across nuanced ethical AI dimensions.

Following model training, we applied the fine-tuned models to the full dataset of 108,998 reviews to infer sentiment for each ethical AI dimension. The distribution of these large-scale predictions is summarized in Table 2.

**Table 3. Fine-Tuning RoBERTa Models. Performance Overview**

| UX Dimension/ Metric | F1 score | Accuracy |
|---|---|---|
| Human Agency & Oversight | 0.74 | 0.75 |
| Diversity & Fairness | 0.91 | 0.91 |
| Societal & Env. Well-Being | 0.92 | 0.92 |
| Accountability | 0.90 | 0.90 |
| Technical Robustness & Safety | 0.74 | 0.74 |
| Privacy & Data Governance | 0.89 | 0.89 |
| Transparency | 0.82 | 0.82 |

Table 3: Model performance overview. *F1 Score* is the weighted average f1 score. Accuracy is the share of predictions in the test-set classified correctly.

### 3.4. Categorizing Product Type and Professional Background

In addition to analyzing sentiment on the seven ethical AI dimensions, we also categorized the type of AI product being reviewed and the professional background of the reviewer. These contextual variables were used to test hypotheses H2 through H4.

To classify product type, we relied on the category metadata available for each product on G2. Based on our theoretical distinction between AI-based end-user applications and developer-oriented platforms, we labeled products as *AI development platforms* if they fell under either the "Data Science and Machine Learning Platforms" or "MLOps Platforms" categories. These categories include tools primarily used to train, deploy, and monitor AI models, and are thus distinct from applications delivering AI-driven outcomes directly to end users.

Similarly, we inferred the professional background of reviewers from their self-reported job titles. Reviewers were classified as having a *technical role* if their job title included keywords such as *engineer*, *developer*, *technical*, or *data scientist*. These roles reflect positions with direct responsibility over the configuration or development of AI systems and are theoretically distinct from non-technical users focused on business or operational outcomes.

## 4. Results

### 4.1 Ethical AI Sentiment and User Satisfaction

To examine how individual ethical AI dimensions relate to user satisfaction, we regressed user star ratings on the sentiment classification for each of the seven ethical AI principles. Reviewer satisfaction is measured using a 1 to 5 star scale, while sentiment on ethical AI dimensions is treated as a discrete variable coded as –1 (negative), 0 (not discussed), or +1 (positive). Thus, the estimated coefficients can be interpreted as the average change in star rating when a given dimension is discussed positively, compared to when it is not discussed (0). Conversely, negative sentiment would be associated with a drop in satisfaction of the same magnitude.
As shown in Table 4, all seven ethical AI dimensions are significantly ($p < .001$) and positively associated with user satisfaction, indicating that ethical alignment, when perceived, is meaningfully tied to how users evaluate AI products overall.

Among the human-oriented dimensions, *accountability* shows the strongest effect: a positive mention of accountability is associated with an increase of approximately 0.36 stars in overall rating. *Societal and environmental well-being* ($\beta = 0.2052$) and *human agency & oversight* ($\beta = 0.1962$) also show sizable effects, suggesting that users value systems that support responsible outcomes and individual control. *Diversity & fairness*, while still significant, shows a weaker association ($\beta = 0.0706$), possibly reflecting its more indirect connection to immediate product experience.

The system-level dimensions show similarly consistent effects. *Technical robustness and safety* ($\beta = 0.1935$) and *transparency* ($\beta = 0.1667$) are both strongly associated with higher satisfaction, emphasizing the importance of reliability and clarity in technical systems. *Privacy and data governance* also shows a significant but slightly more modest effect ($\beta = 0.1163$).

Overall, these results offer strong support for the hypothesis that ethical AI principles – when positively experienced – enhance user satisfaction, across both system-level and human-oriented dimensions. The effect sizes further suggest that these are not trivial associations: a single positive ethical dimension can boost satisfaction by 0.1 to 0.36 stars in overall rating.

Table 4. Ethical AI & User Satisfaction
Dep. Variable: Reviewer Rating

| | Human-Oriented | | | | System-Oriented | | |
|---|---|---|---|---|---|---|---|
| Human Agency & Oversight | 0.1962*** (0.0042) | | | | | | |
| Diversity & Fairness | | 0.0706*** (0.0074) | | | | | |
| Societal & Env. Well-Being | | | 0.2052*** (0.0108) | | | | |
| Accountability | | | | 0.3620*** (0.0066) | | | |
| Technical Rob. & Safety | | | | | 0.1935*** (0.0026) | | |
| Privacy & Data Governance | | | | | | 0.1163*** (0.0075) | |
| Transparency | | | | | | | 0.1667*** (0.0048) |
| Controls | Yes | Yes | Yes | Yes | Yes | Yes | Yes |
| Observations | 103723 | 103723 | 103723 | 103723 | 103723 | 103723 | 103723 |
| R-squared | 0.07 | 0.05 | 0.05 | 0.07 | 0.10 | 0.05 | 0.06 |

Robust Standard errors in parentheses.* p<.1, ** p<.05, ***p<.01. Controls include: Dummies for product category, company age, and number of employees.

### 4.2. Discussion of Ethical AI Dimensions, Product Type and Job Role

In the next step, we examine how the frequency with which users engage with ethical AI dimensions varies by product type (AI developer platforms vs. end-user applications) and professional background (technical vs. non-technical users). Rather than focusing on sentiment, we analyze discussion levels – that is, whether a given ethical dimension is mentioned at all, regardless of whether it is evaluated positively or negatively. This approach allows us to capture salience: which ethical issues are top of mind for different user groups and product types.

To analyze these patterns, we perform Chi-square ($\chi^2$) tests of independence, a widely used method for examining associations between categorical variables in social science and HCI research (Agresti, 2018). Given that both our dependent variables (discussion: yes/no) and independent variables (job role, product type) are categorical, the Chi-square test provides an appropriate and interpretable approach for detecting statistically significant differences in discussion frequencies across groups (McHugh, 2013).

We begin with user role. A Chi-square test on the joint distribution of all ethical dimensions indicates a statistically significant association between user type and discussion frequency ($\chi^2$ = 118.43, p < .001). In other words, whether and how ethical AI dimensions are discussed varies meaningfully between technical and non-technical users. Figure 2 breaks this effect down by individual dimension, showing which topics are more likely to be raised by each group.

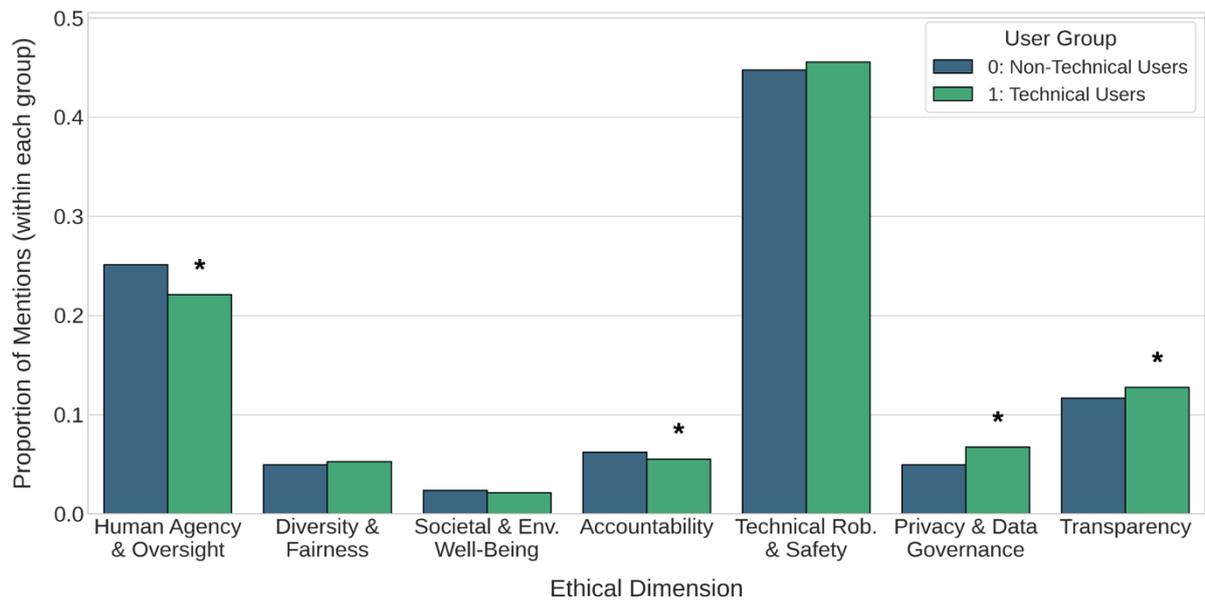

**Figure 2.** Proportion of Discussed Ethical Dimensions by User Group.
(* = statistically significant at p<0.05)

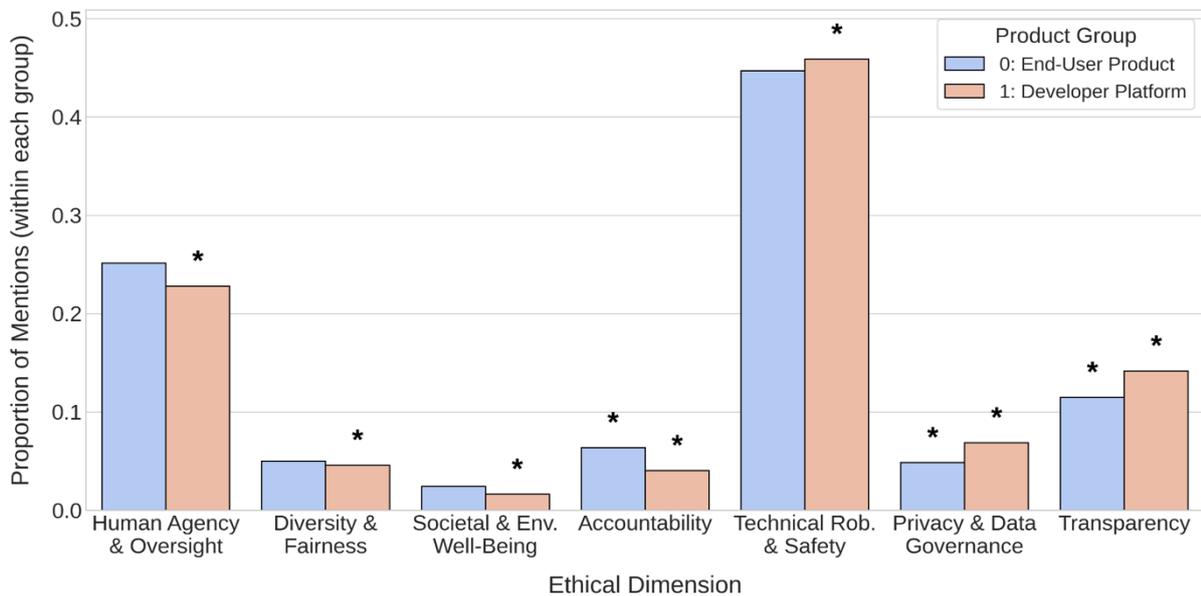

**Figure 3.** Proportion of Discussed Ethical Dimensions by Product Type.
(* = statistically significant at p<0.05)

Overall, we find partial support for H2a: among the four human-oriented ethical dimensions, all except *diversity & fairness* are more frequently mentioned by non-technical users. Two of these differences – *human agency & oversight* and *accountability* – are statistically significant. Conversely, as hypothesized in H2a, system-oriented dimensions are more frequently discussed by technical users. This difference is statistically significant for *privacy & data governance* and *transparency*, but not for *technical robustness & safety*, which is discussed at similar rates by both user groups.

We observe a similar pattern when comparing AI development platforms and end-user applications. As indicated by the Chi-square test ($\chi^2 = 437.05$, $p < .001$), the overall distribution of ethical AI discussions differs significantly by product type. Figure 3 breaks this down by dimension, revealing clear contrasts in which ethical concerns are most salient across the two product categories.

Consistent with H2b, human-oriented dimensions – such as *human agency & oversight*, *accountability*, and *societal & environmental well-being* – are discussed more frequently in reviews of end-user applications. This supports the idea that users of end-user products engage directly with AI outputs and are therefore more sensitive to how the technology aligns with personal agency, fairness, and societal values. The difference is statistically significant for all four dimensions.

Conversely, system-oriented dimensions are more salient in evaluations of AI development platforms. All three system-oriented dimensions are significantly more likely to be discussed in this in reviews on AI development platforms compared to end-user applications.

Together, these findings offer strong support for H2b and reinforce the idea that ethical AI concerns are not uniformly distributed but shaped by the context of use.

### 4.3. Moderating Effect of Product Type and Job Role on the Ethical AI-Satisfaction Link

Table 5. Moderators of the Ethical AI-Satisfaction Link
Dep. Variable: Reviewer Rating

| Ethical AI Dim. | Human-Oriented | | | | System-Oriented | | |
|---|---|---|---|---|---|---|---|
| | Oversight | Diversity | Societal | Account. | Tech. Rob. | Privacy | Transpar. |
| Ethical AI Dim. | 0.3044*** | 0.1055*** | 0.3345*** | 0.5488*** | 0.3035*** | 0.1820*** | 0.2599*** |
| | (0.0063) | (0.0115) | (0.0165) | (0.0100) | (0.0039) | (0.0118) | (0.0076) |
| Technical Role | 0.0809*** | 0.0176 | 0.0261* | 0.0145 | 0.0563*** | 0.0148 | 0.0274** |
| | (0.0168) | (0.0138) | (0.0138) | (0.0137) | (0.0142) | (0.0139) | (0.0139) |
| Dev. Platform | 0.0522*** | 0.0248** | 0.0309*** | 0.0083 | 0.0515*** | 0.0229* | 0.0259** |
| | (0.0143) | (0.0119) | (0.0119) | (0.0117) | (0.0119) | (0.0119) | (0.0119) |
| Technical Role x Ethical AI Dim. | -0.1553*** | -0.0894** | -0.1798*** | -0.2734*** | -0.1554*** | -0.1102*** | -0.1391*** |
| | (0.0263) | (0.0445) | (0.0697) | (0.0431) | (0.0156) | (0.0398) | (0.0288) |
| Dev. Platform x Ethical AI Dim. | -0.0855*** | -0.0961** | -0.1121* | -0.2082*** | -0.0846*** | -0.0622* | -0.0424* |
| | (0.0213) | (0.0389) | (0.0644) | (0.0405) | (0.0123) | (0.0326) | (0.0225) |
| Controls | Yes | Yes | Yes | Yes | Yes | Yes | Yes |
| Observations | 103723 | 103723 | 103723 | 103723 | 103723 | 103723 | 103723 |
| R-squared | 0.04 | 0.02 | 0.02 | 0.05 | 0.08 | 0.02 | 0.03 |

Robust Standard errors in parentheses.* p<.1, ** p<.05, ***p<.01. Controls include: Company age, and number of employees.

Table 5 presents the results of our moderation analysis, designed to test Hypotheses H3 and H4. In all models, the dependent variable is the overall user rating (on a 1–5 star scale), and each column corresponds to a different ethical AI dimension – listed in the column headers. For each dimension, we estimate how its sentiment (positive, negative, or not discussed) relates to satisfaction, and how this relationship is moderated by user role (technical vs. non-technical) and product type (developer platform vs. end-user application).

For example, column 1 focuses on the dimension *human agency & oversight*. The coefficient in the first row shows the baseline effect of sentiment on this dimension on overall satisfaction. The interaction terms further test whether this relationship is stronger or weaker depending on the reviewer's job role or whether the product is a developer tool. Each subsequent column repeats this structure for a different ethical AI principle, allowing us to assess whether moderation effects vary across dimensions.

We find consistent support for Hypotheses H4a and H4b: across all seven ethical AI dimensions, the interaction effects with both technical user role and AI development platform are negative and statistically significant. This indicates that the positive relationship between ethical AI sentiment and user satisfaction is systematically weaker for technical users and for developer-oriented products.

Correspondingly, and contrary to Hypotheses H3a and H3b, we do not observe stronger associations between system-oriented dimensions (e.g., technical robustness, transparency) and satisfaction among technical users or developer platforms. Instead, the moderation effects are broadly negative across all dimensions, suggesting a general attenuation rather than dimension-specific amplification. This provides no empirical support for the differentiated pattern predicted in H3a and H3b.

## 5. Discussion

### 5.1 Ethical AI and User Satisfaction

This study provides strong empirical support for hypothesis H1, which posited that ethical AI principles are positively associated with user satisfaction. Across all seven dimensions of the EU Ethics Guidelines for Trustworthy AI, positive sentiment corresponded to higher satisfaction ratings. This finding affirms that ethical AI is not merely a matter of policy compliance or organizational responsibility; it is also a key driver of how users perceive the quality and value of AI systems. In this sense, ethical alignment functions not only as a governance benchmark but as a central component of product experience, with direct implications for trust, adoption, and satisfaction.

Importantly, the results suggest that ethical AI should be treated as a user experience concern. Existing work in Human-AI Interaction (HAI) and Human-Computer Interaction (HCI) emphasizes the role of interpretability, robustness, and usability in supporting trust and acceptance (Amershi et al., 2019; Shneiderman, 2020). Our findings extend this literature by demonstrating that a broader set of ethical principles, including fairness, accountability, human agency, and societal well-being, also meaningfully contribute to user satisfaction. These principles are not abstract or peripheral. They operate as experiential affordances that shape users' emotional responses, sense of control, and evaluations of legitimacy.

This user-facing role of ethics supports a broader reframing of AI ethics as a design concern. Rather than conceptualizing ethical alignment as a back-end compliance issue that remains invisible to end users, product teams should treat ethical dimensions as elements of the interface and user experience. Just as responsiveness, visual design, or usability are surfaced through deliberate interaction design, so too should dimensions like transparency, accountability, and user control be actively communicated. This requires operationalizing ethics in ways that are perceptible and usable, for example, through redress mechanisms, explainability interfaces, and consent flows.

While all ethical dimensions showed positive associations with satisfaction, not all were equally prominent in user discourse. For example, dimensions such as *Societal and Environmental Well-being*, *Diversity and Fairness*, and *Privacy and Data Governance* were each discussed in fewer than 10% of reviews. This suggests that users may not spontaneously articulate these values unless they are made salient. However, when these dimensions were mentioned, they had a substantial positive association with satisfaction. These findings point to a gap between the ethical features users care about and those that are currently visible or foregrounded in AI systems. Designers and developers should therefore consider how to more effectively surface these underrepresented values within product interfaces and messaging.

In sum, ethical AI is not an abstract add-on or regulatory afterthought. It is embedded in how users experience, evaluate, and ultimately trust AI systems. When ethical principles are implemented in ways that are both functional and perceptible, they significantly enhance user satisfaction. These results call for a shift in orientation: from ethics as external oversight to ethics as an integral part of user experience.

### 5.2. Discussion of Ethical Dimensions by User Role and Product Type

The analysis provides strong support for the proposition that the salience of ethical AI dimensions varies systematically by user role and product type, affirming H2a and H2b. Technical users – those more likely to engage with AI systems at the level of infrastructure, model development, or deployment – more frequently emphasized system-level ethical principles, including technical robustness, transparency, and data governance. By contrast, non-technical users, who typically interact with AI through its outputs (e.g., recommendations, content, or decisions), placed greater emphasis on human-oriented dimensions such as fairness, accountability, and human agency. A similar pattern emerged across product types: reviews of developer-facing tools (e.g., MLOps and data science platforms) highlighted system-centric concerns, while feedback on end-user applications more often reflected normative, interactional, or societal considerations.

These patterns are consistent with the idea that ethical AI is not perceived in a vacuum; it is shaped by a user's position in the AI lifecycle. System-facing users work directly with the inner workings of AI models and infrastructure; they are responsible for configuring, tuning, and debugging AI systems, and thus their ethical concerns are oriented toward system performance, explainability, and reliability. Outcome-facing users, in contrast, engage with AI systems primarily through their effects – what the AI outputs, recommends, or decides – and are thus more attuned to whether the system behaves fairly, respects autonomy, or aligns with societal norms. This lifecycle-based distinction helps explain why different ethical principles are emphasized by different groups, even when interacting with the same underlying technologies.

Moreover, the analysis reveals that some dimensions, despite being normatively significant, are less frequently discussed by technical users or in developer-platform reviews. For instance, fairness, societal well-being, and human agency appear far less often in system-facing contexts, even though they are key pillars of ethical AI. This suggests that such values may be underrepresented in system workflows – not because they are unimportant, but because they are less visible or actionable during development. These findings point to a critical design challenge: how to surface and operationalize underrepresented ethical concerns in environments where they are not naturally prominent.

Taken together, the results highlight the need for a more context-aware approach to ethical AI. If ethical principles are to be meaningful in practice, they must be embedded in ways that reflect users' actual roles and relationships to AI systems. For developers and system-facing users, this may involve tools for bias auditing, fairness diagnostics, or transparency-by-design in model pipelines. For outcome-facing users, ethical AI must be made tangible through interfaces, content framing, and redress mechanisms that allow users to understand and contest decisions. In this sense, ethical design is not a single intervention but a distributed responsibility – one that must account for where the user sits in the AI lifecycle and what aspects of the system they are empowered to control.

### 5.3. Differential Impacts of Ethical AI: The Moderating Role of User and Product Contexts

While prior sections established that user background and product type influence which ethical principles users emphasize, our moderation analysis reveals a more complex relationship between these factors and satisfaction. Specifically, we do not find support for H3a and H3b: system-level ethical dimensions are not more strongly associated with satisfaction among technical users or in reviews of developer platforms. Instead, support is found for H4a and H4b: ethical dimensions are more strongly associated with satisfaction for non-technical users and end-user applications for all dimensions.

This pattern suggests that the influence of ethical principles on satisfaction is shaped less by topical alignment – that is, the ethical concerns most frequently mentioned – and more by users' perceived control and vulnerability. Technical users and system-facing reviewers may frequently discuss issues like transparency or robustness, but these concerns appear to exert less weight on their satisfaction judgments. This may reflect a greater sense of agency: when ethical shortcomings arise, technical users are more likely to view them as manageable or fixable within their development workflows. By contrast, non-technical users typically experience AI systems through their outputs and lack the tools to diagnose or remedy problems, making ethical flaws more impactful.

These results have important implications: First, they highlight the importance of increasing the visibility and intelligibility of system-level ethical dimensions within user interfaces; particularly for non-technical users. While these users rarely mention technical features such as robustness, data governance, or transparency, our results show that when these dimensions are recognized, they have a strong positive influence on satisfaction – stronger than for technical users. Similarly, users of AI-based end-user products (e.g., chatbots, marketing assistants, HR tools) are more strongly affected by technical flaws, yet often lack the system-level visibility that developer tools naturally provide. Designers should therefore prioritize surfacing key system-level features in accessible ways – for example, through confidence indicators, warnings about instability, or clear explanations of data use. Improving the intelligibility of technical aspects can support trust and decision-making even among users without deep technical knowledge.

Second, the results point to a broader gap between ethical implementation and ethical experience. Many ethical AI safeguards, such as fairness audits, documentation standards, or internal governance processes, are implemented on the back end, largely invisible to those interacting with the final product. While these interventions are critical for developers and platform-level governance, they may fail to impact users of consumer-facing AI tools if not translated into concrete, user-facing features. In these contexts, ethical principles must be operationalized in ways that are not only technically rigorous but experientially meaningful; integrated directly into the touchpoints where users interact with AI systems.

Third, ethical risks tend to concentrate downstream – where users have the least visibility, understanding, and control over how AI works. Non-technical users, as well as those engaging with AI through end-user applications, encounter the system through its outputs rather than its internals. They are less able to interrogate or contest decisions, yet more exposed to the real-world consequences of ethical shortcomings – whether biased recommendations, lack of transparency, or brittle functionality. These findings call for stronger protections at the point of use, including contestability mechanisms, feedback loops, and human-in-the-loop options in settings where users face high stakes and low agency.

Finally, the results reinforce the need to include non-expert, end-user perspectives in the design and governance of ethical AI. Current frameworks remain predominantly top-down, shaped by technical experts, legal scholars, and policymakers. However, our findings suggest that users who lack technical knowledge – and especially those engaging with AI through outcome-oriented products – are both more vulnerable to ethical failures and more frequently excluded from governance conversations. Incorporating their perspectives through participatory design processes, feedback channels, and user-centered evaluation practices can help identify blind spots and ensure that ethical principles align with actual user needs and risks.

**5.4. Limitations & Future Research**

This study offers one of the first large-scale empirical assessments of how ethical AI dimensions shape user satisfaction across real-world systems. At the same time, several limitations are worth noting that suggest fruitful directions for future research.

One limitation lies in the use of product reviews as the primary data source. While this approach provides scalable, naturalistic access to user evaluations across a wide range of AI products, it also introduces important biases. Users who leave public reviews may differ systematically from the broader user base; for instance, in motivation, experience level, or expectations. Moreover, the platform's user base may overrepresent certain roles, industries, or organizational contexts, limiting generalizability. Future research could address these limitations by incorporating more representative sampling methods or combining review data with structured user panels.

A second methodological limitation concerns the exclusive reliance on textual data to capture perceptions of ethical AI. While user reviews offer valuable, unprompted expressions of evaluative sentiment, they inherently limit the depth and dimensionality of user experience. Our use of sentiment analysis provides a scalable proxy for satisfaction but may miss ambivalence, nuance, or mixed sentiments, particularly in ethically complex domains. Similarly, the detection of ethical AI dimensions depends on the presence of explicit language, meaning that ethical concerns users feel but don't articulate may go unobserved. Future research could complement this approach with survey-based instruments, multi-dimensional satisfaction scales, or qualitative methods such as interviews or think-aloud protocols to more fully capture how users recognize and evaluate ethical features.

Another limitation relates to the operationalization of ethical AI through the dimensions outlined in the EU Ethics Guidelines for Trustworthy AI. We believe this is a conceptually sound choice, as these dimensions encompass many of the key concerns identified in the academic literature on ethical AI (Papagiannidis et al., 2023; Floridi & Cowls, 2022). Moreover, other widely used frameworks – such as the OECD AI Principles, which emphasize transparency, robustness, accountability, human-centered values, and sustainable development – overlap substantially with the EU dimensions. Nonetheless, ethical priorities can vary across sectors, cultures, and stakeholder groups. Relying on a single framework risks overlooking alternative value systems or context-specific concerns. Given the complexity of our methodological pipeline, comparing multiple frameworks was beyond the scope of this study. However, future research could explore how ethical perceptions shift under different normative models or regulatory regimes, or examine the salience of specific ethical dimensions in domains such as healthcare, education, or criminal justice.

Looking ahead, several additional theoretical and practical extensions are possible. Future research could investigate how users come to recognize ethical issues in the first place; whether through specific interface cues, contextual signals, or social amplification. Experimental and longitudinal methods could help identify which types of design interventions increase ethical awareness, trust, or contestability. Additionally, integrating end-user perspectives into the development of ethical AI guidelines remains an important frontier. Current governance models are often expert-led and top-down; incorporating structured, bottom-up feedback from those who interact with AI outputs, particularly non-technical users, could help make ethical frameworks more relevant, inclusive, and grounded in actual use.

## 6. Conclusion

This study provides large-scale empirical evidence that ethical AI principles meaningfully shape how users evaluate AI systems in real-world settings. Across a wide range of products and roles, we find that ethical dimensions such as transparency, accountability, and fairness are not only noticed by users but significantly influence their satisfaction; especially when those users are further removed from development and control. These findings reframe ethical AI as not merely a compliance issue, but as a core component of user experience and product quality. They also highlight the importance of making ethical principles more visible and accessible to end users, and of incorporating diverse user perspectives into ethical AI governance. As AI systems continue to mediate critical decisions and services, aligning their design with both ethical standards and user expectations will be essential to fostering trust, usability, and long-term acceptance.

**Declaration of Conflicting Interests**
The authors certify that no potential conflicts of interest with respect to the research, authorship, and/or publication of this article.

**Acknowledgment**
This research was supported by Culture, Sports and Tourism R&D Program through the Korea Creative Content Agency grant funded by the Ministry of Culture, Sports and Tourism in 2024(Project Name: Development of multimodal UX evaluation platform technology for XR spatial responsive content optimization, Project Number: RS-2024-00361757). This work was supported by Hankuk University of Foreign Studies Research Fund (of 2025)